\documentclass[manuscript]{aastex}
\usepackage{amssymb}
\usepackage{ctable}
\usepackage{morefloats}
\usepackage[colorlinks,breaklinks,citecolor=blue]{hyperref}
\usepackage{graphicx,epsfig,fancyhdr,psfig,rotating,amsmath,epsf,txfonts,natbib,epstopdf,multirow,appendix}

\def \kms {{ \rm km\;s$^{-1}$}}

\def \arcsec {$^{''}$}

\graphicspath{{figs/}}
\begin{document}
\title{Sources of quasi-periodic propagating disturbances above a solar polar coronal hole}
\author{ Fangran Jiao\altaffilmark{1}, Lidong Xia\altaffilmark{1}, Bo Li\altaffilmark{1}, Zhenghua Huang\altaffilmark{1}, Xing Li\altaffilmark{2},
Kalugodu Chandrashekhar\altaffilmark{1}, Chaozhou Mou\altaffilmark{1}, Hui Fu\altaffilmark{1}}

\altaffiltext{1}{Shandong Provincial Key Laboratory of Optical Astronomy and Solar-Terrestrial Environment, Institute of Space Sciences, Shandong University, Weihai, 264209 Shandong, China \textit{xld@sdu.edu.cn}}
\altaffiltext{2}{Department of Physics, Aberystwyth University, Aberystwyth SY23 3BZ, UK}

\date{Received date, accepted date}

\begin{abstract}
Quasi-periodic propagating disturbances (PDs) are ubiquitous in polar coronal holes on the Sun. It remains unclear as to what generates PDs. In this work, we investigate how the PDs are generated in the solar atmosphere by analyzing a four-hour dataset taken by the Atmospheric Imaging Assembly (AIA) on board the Solar Dynamics Observatory (SDO). We find convincing evidence that spicular activities in the solar transition region as seen in the AIA 304\,\AA\ passband are responsible for PDs in the corona as revealed in the AIA 171\,\AA\ images. We conclude that spicules are an important source that triggers coronal PDs.
\end{abstract}

\keywords{Sun: atmosphere - Sun: chromosphere - Sun: transition region - Sun: corona}

\maketitle

\section{Introduction}
\label{sect_intro}
Quasi-periodic propagating disturbances (PDs) in solar polar coronal holes have been reported since the \mbox{\textit{Skylab}} era.
When analyzing the Mg\,{\sc x}\,625\,\AA\ line radiance in plumes, \citet{1983SoPh...89...77W} was the first
  to discover intensity disturbances with some 10\% variation and propagating speeds of 100--200\,\kms.
Since the launch of SoHO, this phenomenon has been extensively studied.
With observations taken off-limb at 1.01--1.2\,R$_\odot$ by the Extreme ultraviolet Imaging Telescope,
  \citet{1998ApJ...501L.217D} found PDs with 10--20\% intensity variations, quasi-periodicities of 10--15\,mins and propagating speeds of 75--150\,\kms\ in fine structures (with a width of 3\arcsec--5\arcsec) of polar plumes.
By using high spatial resolution data taken by the Atmospheric Imaging Assembly (AIA) on board the Solar Dynamics Observatory (SDO),
  PDs are found in both plume and inter-plume regions.
In the AIA data, the periodicities of PDs are found to be 10--30 mins, and the propagating speeds are 100--170\,\kms\,\citep{2011A&A...528L...4K}.
Spectral measurements of plumes made with, e.g., CDS and SUMER on board SoHO, and EIS on board Hinode,
   also showed the existence of PDs with periodicities of 10--30 mins and relative intensity variations of a few percent\,\citep{2000SoPh..196...63B,2009A&A...499L..29B}.
Furthermore, H\,{\sc i} Ly$\alpha$ measurements with UVCS/SoHO indicate that signatures of PDs are likely to be present
   up to 2.2~$R_\odot$\,\citep{2004ApJ...605..521M}.

\par
The fact that PDs tend to propagate with a speed close to the sound speed suggests that they are likely to be signatures
   of propagating slow magnetic-acoustic waves\,\citep{1998ApJ...501L.217D,2000ApJ...533.1071O,2012A&A...546A..93G}.
Alfv\'en waves and fast magnetic-acoustic waves were also suggested to produce PDs in inter-plume regions\,\citep{2010ApJ...718...11G}.
On the other hand, by analyzing EIS/Hinode and AIA/SDO data, \citet{2009ApJ...701L...1D} revealed the presence of quasi-periodic outflows
   in the footpoints of coronal loops.
Furthermore, \citet{2011ApJ...736..130T} discovered that high-speed quasi-periodic outflows can be found in both plume and inter-plume regions,
   and the outflow speeds are similar to the propagating speeds of PDs.
These outflows have been suggested to be sources of PDs\,\citep{2010A&A...510L...2M}.

\begin{figure*}[!ht]
\includegraphics[width=17cm,clip,trim=1.5cm 2.8cm 2.5cm 6cm]{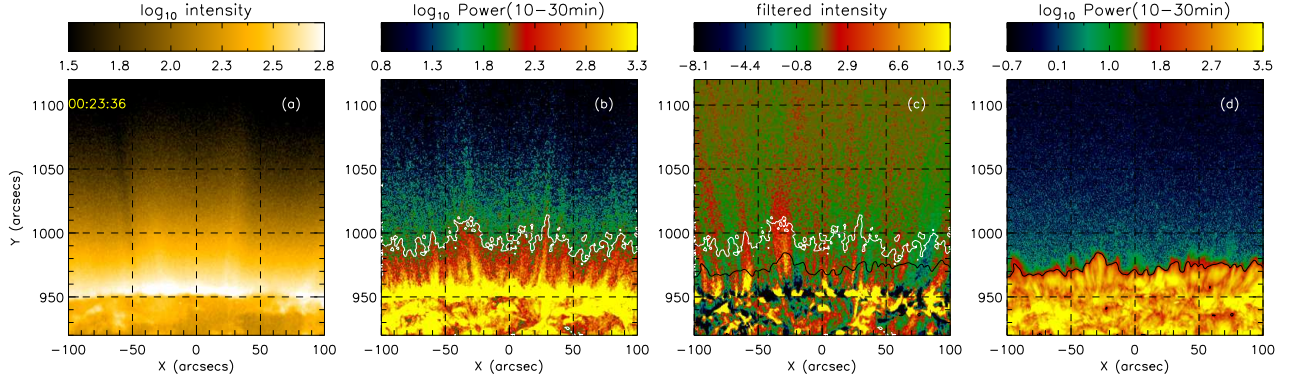}
\caption{Field-of-view studied in the present work. Panel (a) gives the original AIA 171\,\AA\ image, (b) shows the spatial distribution of wavelet power summed over periodicities of 10--30 mins, (c) presents the band-pass (10--30 mins) filtered images, and (d) displays the wavelet power summed over periodicities of 10--30 mins for the AIA 304\,\AA. The white contours over-plotted on (b) and (c) outline the structures of PDs determined from (b), while the black lines (panels c and d) represent the contour of the AIA 304\,\AA\ power image (i.e. panel d).}\label{fig1}
\end{figure*}

\begin{figure*}[!ht]
\includegraphics[width=17cm,clip,trim=0cm 11.5cm 0cm 0cm]{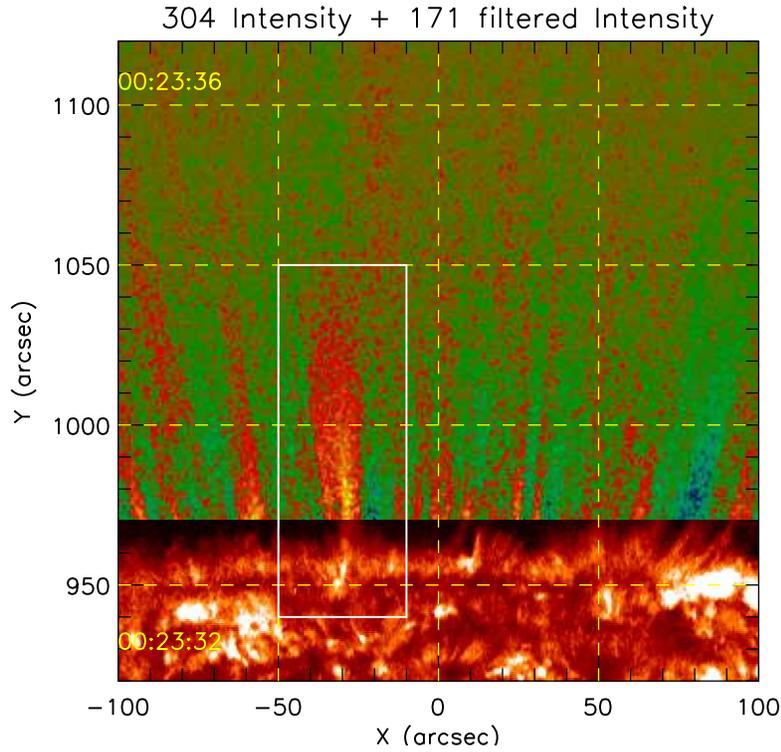}
\caption{Composite image constructed from the original AIA 304\,\AA\ (bottom) and filtered AIA 171\,\AA\ (top) images. Spicules are visible on the AIA 304\,\AA\ image, while PDs are clear in the filtered AIA 171\,\AA\ images. The white solid lines denote the region that is studied in detail in Figs.~\ref{fig3}, \ref{fig4} and \ref{fig5}. The accompanying on-line animation makes it easy for one to follow the evolution of the spicular activities and the PDs.\label{fig2}}
\end{figure*}

\begin{figure*}[!ht]
\includegraphics[width=17cm,clip,trim=1cm 5cm 2cm 2cm]{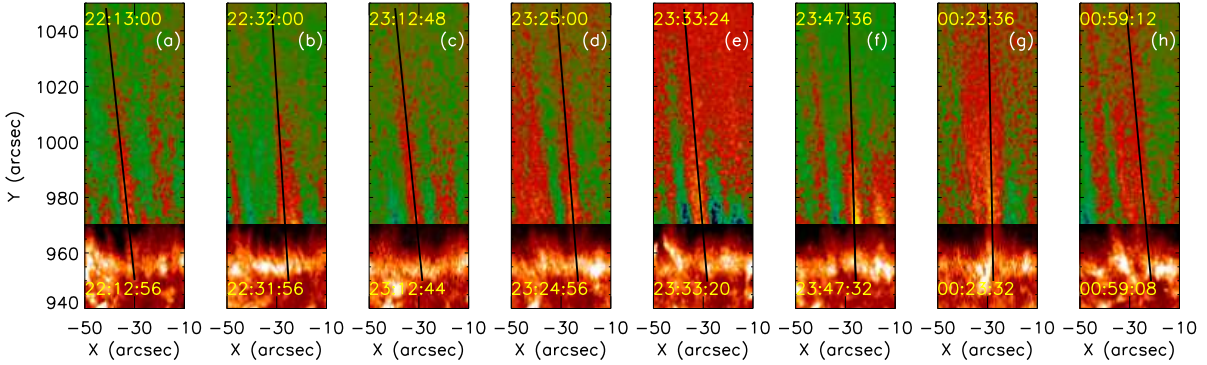}
\caption{Snapshots showing the connection between spicules in the AIA 304\,\AA\ images (bottom) and PDs in the filtered AIA 171\,\AA\ images (top). The solid line on each image denotes the cut used to produce the time-slice plots in Fig.~\ref{fig4}. An animation is given online. \label{fig3}}
\end{figure*}

\begin{figure*}[!ht]
\includegraphics[width=17cm,clip,trim=1cm 5cm 2cm 2cm]{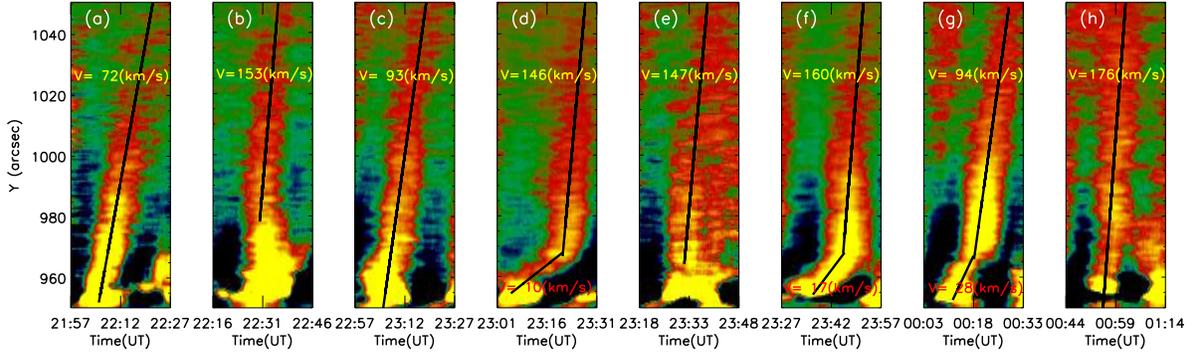}
\caption{Time-slice images of the PDs shown in Fig.~\ref{fig3} (here only AIA 171\,\AA\ measurements are used). The solid lines are the slopes from which the speeds of the propagating structures are determined. \label{fig4}}
\end{figure*}

\begin{figure*}[!ht]
\includegraphics[width=17cm,clip,trim=0cm 0.3cm 0cm 0cm]{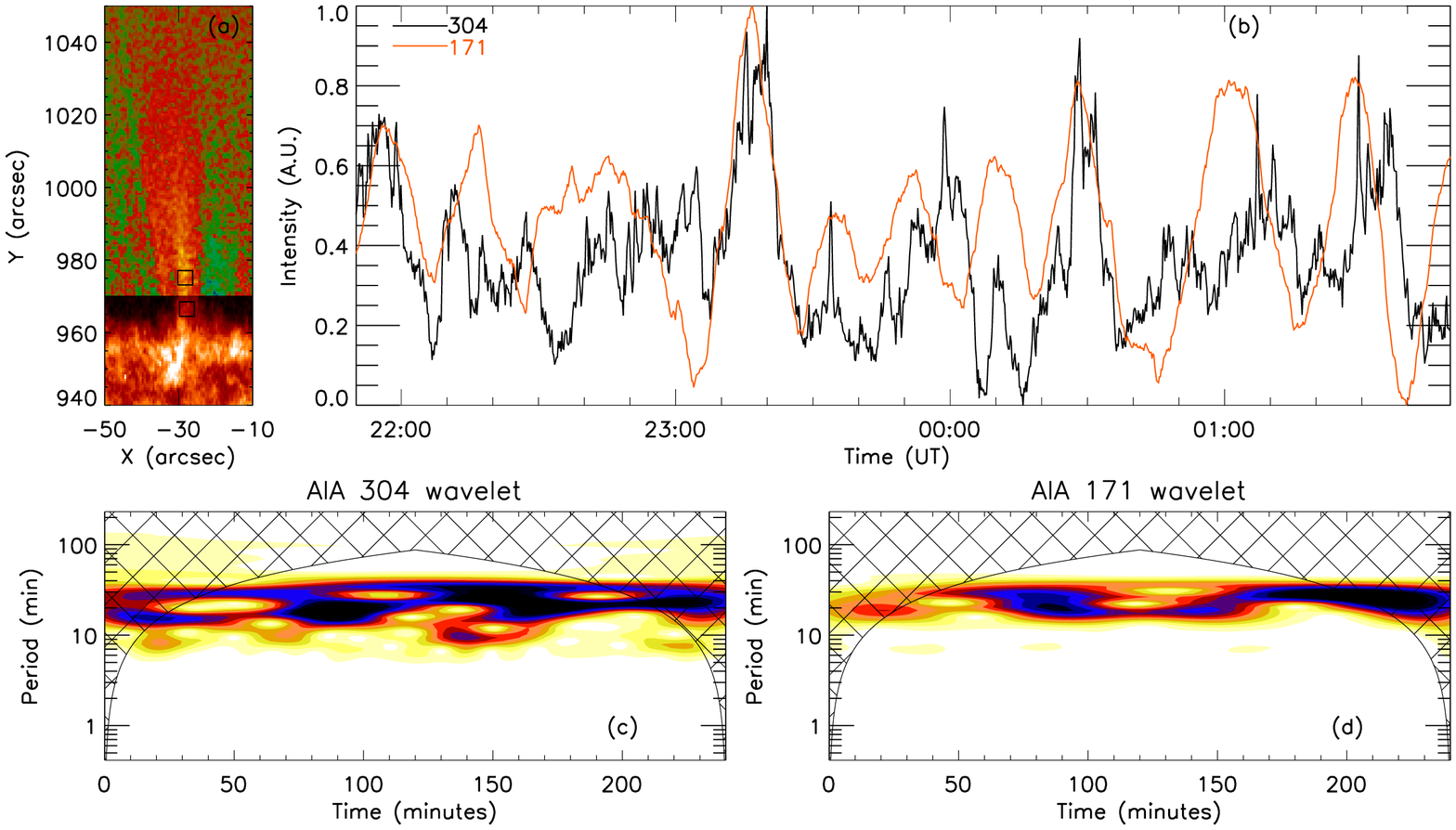}
\caption{Lightcurves of the AIA\,304\,\AA\ (b, the black line) and the filtered AIA 171\,\AA\ (b, the red line) taken from two small boxed regions  marked on (a) (black solid lines). The wavelet power spectra of the two lightcurves are also given (c: original 304\,\AA, d: filtered 171\,\AA).\label{fig5}}
\end{figure*}

\par
How PDs are generated remains debated.
Although an obvious candidate source, magneto-acoustic-gravity (MAG) waves in the photosphere with periods longer than $\sim 10$~mins
   cannot readily penetrate the chromosphere and make it into the corona, since the periods are substantially longer
   than some critical value $P_{\rm c}$ \citep{1977A&A....55..239B}.
For a low-beta environment, this $P_{\rm c}$ reads $4\pi c_s/(\gamma g \cos\theta)$ where $\gamma=5/3$ is the adiabatic index,
   $g = 27400$~cm~s$^{-2}$ is the Sun's gravitational acceleration, $c_s$ is the sound speed,
   and $\theta$ is the angle between the wavevector and the background magnetic field \citep{2014A&A...561A..19Y}.
Adopting a $c_s = 7$~km~s$^{-1}$ as being representative of chromospheric values, one finds that $P_{\rm c} = 192/\cos\theta$~secs, meaning that
    MAG waves with periods $\gtrsim 10$~mins can reach the corona only along highly inclined magnetic flux tubes with $\theta \gtrsim 70^\circ$.
However, if PDs have coronal sources, the restriction due to $P_{\rm c}$ is not as stringent.
One possibility of this kind will be that PDs are the coronal response to spicules
    impinging on the corona in a repeated manner.
Indeed, SUMER measurements show that spicules, the omnipresent ejections of cool chromospheric plasmas to coronal heights,
    can occur repeatedly at the same place\,\citep{2005A&A...438.1115X} with periods of 10--15 mins\,\citep{2005ESASP.592..575X}.
Another support for this possibility comes from the recent numerical studies by \citet{2012ApJ...754..111O} and \citet{2013ApJ...775L..23W}
    who showed that any upflow pulse at the coronal base, regardless of its origin,
    leads to propagating intensity disturbances guided by magnetic flux tubes.

\par
In the present study, by analyzing the high resolution AIA data we aim to examine whether spicules can be a source of PDs.
This manuscript is organized as follows. Section\,\ref{sect_obs} describes the observational data, method of data analysis and our results.
Section\,\ref{sect_con} summarizes our conclusions.

\section{Observations, data analysis and results}
\label{sect_obs}
The data analyzed in this study were taken from 21:50\,UT, 5 August 2010 to 01:50\,UT, 6 August 2010 by the AIA\,\citep{2012SoPh..275...17L}. The spatial resolution of the data is about 1.2\arcsec and the cadence is 12\,s. The AIA monitors the full disk of the Sun in 10 wavelength bands. In this work, we chose to use the passbands of 304\,\AA\ (mainly contributed by He\,{\sc ii} with response function peaking at 0.05 MK) and 171\,\AA\ (mainly contributed by Fe\,{\sc ix} with response function peaking at 0.8 MK). The AIA 304\,\AA\ passband can well observe spicules in the transition region and the 171\,\AA\ one is commonly used to identify PDs in the corona. We focus our investigation on the north polar region with solar X between $-$100\arcsec and 100\arcsec\ where a coronal hole is clearly seen (Figure\,\ref{fig1}a).

\par
We started with applying a wavelet tool kit to the time series of every pixel of the AIA 171\,\AA\ images.
The wavelet power in a selected range of periodicities was then used to construct a power image.
An example is given in Fig.~\ref{fig1}b, where we show the spatial distribution of wavelet power with periodicities between 10 and 30 mins at 00:23\,UT.
Note that constructing a wavelet power map involves a sequence of images over a certain time interval.
However, it is possible that a power map is assigned a particular time around which this interval is centered.
By following the time series of the power images, we found that PDs are an inherent structure above the polar coronal hole (see the elongated structures in Fig.~\ref{fig1}b). Although most of the PD structures are clearly seen below 1.1\,R$_\odot$ (i.e. solar Y being 1050\arcsec), some PDs show clear structures reaching an altitude of 1.15\,R$_\odot$ (i.e. solar Y being 1100\arcsec) or even higher. They are clearly present in power images at periodicities from 10 mins to 30 mins, which are consistent with previous studies\,\citep{1998ApJ...501L.217D,2000SoPh..196...63B,2009A&A...499L..29B,2011A&A...528L...4K}. To compare the PDs with spicules that were clearly visible in the AIA 304\,\AA\ channel, we further applied a simple band-pass filter to the AIA 171\,\AA\ images. The time series of intensity at each pixel were smoothed with a 50-point window to filter out signals with periods shorter than 10 mins. The resultant time series were then subtracted by a 150-point running average of the original time series to filter out signals with periods longer than 30 mins. The snapshot at 00:23\,UT of the final product, essentially a band-pass filtered image series, is shown in Fig.~\ref{fig1}c. While Figs.~\ref{fig1}b and \ref{fig1}c look similar, we find that the temporal evolution of the PDs can be better followed in the band-pass filtered images. This is crucial when comparing the PDs with the rapidly-evolving spicular activities.

\par
Figure\,\ref{fig2} and the online animation show composite images constructed from the original AIA 304\,\AA\ images with solar Y ranging from 920\arcsec to 970\arcsec and
   band-pass filtered AIA 171\,\AA\ images with solar Y ranging from 970\arcsec to 1120\arcsec.
The AIA 304 \AA\ channel observes plasmas at around 0.05 MK, capturing the dynamics of the solar transition region and is ideal for identifying spicules.
While the PDs are well identifiable in the AIA 171\,\AA\ filtered images, the series of the composite images is the best approach to
  investigate the correlation between the PDs in the corona and the spicules in the transition region.
The animation clearly demonstrates that,
when a spicule is discernible in the AIA 304\,\AA\ channel, a PD is seen to be generated and/or amplified in the AIA\,171\,\AA.
Note that a one-to-one correspondence between spicules and PDs is not found in some cases, see our discussion in connection to Fig.~\ref{fig5}.

\par
We further examined a small area denoted by the white box in Fig.~\ref{fig2}.
In Fig.~\ref{fig3} and the online animation, we present the evolution of the small area viewed in the composite images. Again, it demonstrates that spicular activities in the solar transition region are a source of the PDs in the corona. We also measured the speeds of the PDs by producing time-slice images along the propagation direction (the solid lines in Fig.~\ref{fig3}). The speeds measured from these PDs vary from 70\,\kms\ to 170\,\kms, which are consistent with previous results\,\citep{1983SoPh...89...77W,1998ApJ...501L.217D,2011A&A...528L...4K}.
In a few cases (see Figs.~\ref{fig4}d, \ref{fig4}f, and \ref{fig4}g), we found that the propagating structures have lower speeds (10--30\kms) near the limb before gaining much higher speed above. We suggest that the speed near the limb represents the speed of the spicular plasmas because the spicules can also be seen near the limb in the AIA 171\,\AA\ channel. The triggered PDs, on the other hand, are propagating with a much higher speed.

\par
The PD speeds measured here are significantly larger than those of classic type I spicules\,\citep[$\sim$20\,\kms,][]{1992str..book.....M} and are close to those of type II spicules\,\citep[50--150\,\kms,][]{2007PASJ...59S.655D}.
One may then ask whether type II spicules are a direct trigger of the PDs?
To address this issue, we further analyzed the upward speeds of spicules that are found to trigger PDs in the corona.
Because spicules viewed in the AIA 304\,\AA\ are too crowded, only 32 relatively isolated examples were selected and their upward speeds were measured. Among these, 17 (15) spicules were found to have upward speeds larger (less) than $50$\,\kms.
Taking a speed of $50$\,\kms\ as the critical value separating type I from type II spicules, one would then conclude that
   these two types of spicules tend to be equally important in triggering PDs.
However, it is not straightforward to identify in the AIA 304~\AA\ images the counterpart of type II spicules, the thin
   elongated structure originally found in Hinode/SOT Ca\,{\sc ii h} emissions.
The reason is twofold.
First, with a bandwidth as wide as 40\,\AA, the AIA 304\,\AA\ passband can actually record emissions
   formed at a variety of temperatures.
Second, the spatial resolution in AIA 304\,\AA\ is $\sim 1.2$\arcsec, substantially less good than
   what is achieved by SOT ($\sim 0.3$\arcsec).
An analysis combined with spectroscopic data will be crucial for associating features in
   the AIA 304\,\AA\ passband with type II spicules.

\par
Our attention has been drawn to a very recent work by \citet{2015ApJ...807...71P}, where
   the authors suggest that small-scale jet-like features with speeds of $\sim 10$\,\kms\ occurring
   in chromospheric network (seen in IRIS 1330\,\AA\ slit-jaw images)
   might be responsible for the PDs in on-disk plumes.
As has already been pointed out by the authors,
   measuring the speeds of those jet-like features is not straightforward
   due to the projection effect.
In this sense, it remains an open question as to whether the jet-like features associated with PDs in on-disk plumes
   are spicules.
Nonetheless, the present study and the work by \citet{2015ApJ...807...71P} suggest
   that PDs in polar coronal holes and PDs in on-disk plumes may both
   have chromospheric sources.
Given the possible difference in the local magnetic field topology in the two regions,
   it may be that this magnetic topology plays an important role in
   determining exactly what in the chromosphere triggers PDs in the corona.
This intriguing question certainly warrants a dedicated study, which is beyond the scope of the present paper.

\par
One may now ask why intensity disturbances triggered by spicules are quasi-periodic.
An obvious possibility is that the spicules themselves possess similar periodicities, e.g., in connection to their quasi-periodic occurrence at the same location (see introduction).
This possibility is examined in Fig.~\ref{fig5}, which illustrates the quasi-periodic nature of the spicular activity and its relation to PDs.  Figure\,\ref{fig5}b shows the lightcurves of the AIA 304\,\AA\ (black line) and the filtered AIA 171\,\AA\ (red line) taken from the lower and higher boxes marked in Fig.~\ref{fig5}a, and Figs.~\ref{fig5}c and \ref{fig5}d show the corresponding wavelet spectra. We remark that the two boxes are selected such that they represent the spicule in the 304\,\AA\ channel and the corresponding PDs in the 171\,\AA\ channel.
At first glance, the two lightcurves are quite well correlated in general (see Fig.~\ref{fig5}b).
However, some spikes in the AIA 304\,\AA\ lightcurve do not have a clear counterpart in the lightcurve of the 171\,\AA.
The reason for this is threefold.
First, the spicules are rather crowded with a variety of directions, making it difficult to distinguish
    their contributions to PDs in the AIA 171\,\AA\ channel.
Second, spicules tend to experience substantial transverse displacement, meaning that the lightcurve gathered from
    one fixed location may correspond to a substantial number of different spicules rather than
    following an individual one.
Third, it is likely that other non-spicular features also contribute
   to these spikes in the AIA 304\,\AA\ lightcurve.
A further investigation via wavelet analyses (Figs.~\ref{fig5}c and \ref{fig5}d) also confirms that the lightcurve of the AIA 304\,\AA\ owns periodicities similar to those in the AIA 171\,\AA\ lightcurve, i.e. 10--30\,mins.
Furthermore, the consistency in the periodicities is found not only in the small region shown in Fig.\,\ref{fig5},
but also in the whole field-of-view studied here.
A power map of AIA 304\,\AA\ summed over periodicities of 10--30 mins is displayed in Fig.\,\ref{fig1}d,
   which correlates well with the PDs shown in AIA 171\,\AA\ images (Figs.\,\ref{fig1}b and \ref{fig1}c).
This again supports the notion that the spicules seen in the transition region are a source of the PDs in the corona.

\section{Conclusion}
\label{sect_con}
In this study, we have analyzed a four-hour dataset of a polar coronal hole observed by the AIA. Propagating disturbances (PDs) with speeds of 70--170\,\kms\ and periodicities of 10--30\,mins are clearly seen in the AIA\,171\,\AA\ images. By following the evolution of the spicular activities in the solar transition region (seen in the AIA 304\,\AA\ channel) and the PDs in the corona (seen in the AIA 171\,\AA\ channel), we found clear evidence that spicules are one of the important sources of PDs.

\acknowledgments
{\it Acknowledgments:}
We thank the anonymous referee for his/her constructive comments. This research is supported by the China 973 program 2012CB825601, the National Natural Science Foundation of China under contracts: 41404135 (ZH), 41274178 and 41474150 (LX \& ZH), and 41174154, 41274176 and 41474149 (BL), the Shandong provincial Natural Science Foundation ZR2014DQ006 (ZH). AIA data is courtesy of SDO (NASA). We thank JSOC for providing downlinks of the SDO data.

{\it Facilities:} \facility{SDO/AIA}.


\begin{thebibliography}{22}
\expandafter\ifx\csname natexlab\endcsname\relax\def\natexlab#1{#1}\fi

\bibitem[{{Banerjee} {et~al.}(2000){Banerjee}, {O'Shea}, \&
  {Doyle}}]{2000SoPh..196...63B}
{Banerjee}, D., {O'Shea}, E., \& {Doyle}, J.~G. 2000, \solphys, 196, 63

\bibitem[{{Banerjee} {et~al.}(2009){Banerjee}, {Teriaca}, {Gupta}, {Imada},
  {Stenborg}, \& {Solanki}}]{2009A&A...499L..29B}
{Banerjee}, D., {Teriaca}, L., {Gupta}, G.~R., {et~al.} 2009, \aap, 499, L29

\bibitem[{{Bel} \& {Leroy}(1977)}]{1977A&A....55..239B}
{Bel}, N. \& {Leroy}, B. 1977, \aap, 55, 239

\bibitem[{{De Pontieu} {et~al.}(2007){De Pontieu}, {McIntosh}, {Hansteen},
  {Carlsson}, {Schrijver}, {Tarbell}, {Title}, {Shine}, {Suematsu}, {Tsuneta},
  {Katsukawa}, {Ichimoto}, {Shimizu}, \& {Nagata}}]{2007PASJ...59S.655D}
{De Pontieu}, B., {McIntosh}, S., {Hansteen}, V.~H., {et~al.} 2007, \pasj, 59,
  655

\bibitem[{{De Pontieu} {et~al.}(2009){De Pontieu}, {McIntosh}, {Hansteen}, \&
  {Schrijver}}]{2009ApJ...701L...1D}
{De Pontieu}, B., {McIntosh}, S.~W., {Hansteen}, V.~H., \& {Schrijver}, C.~J.
  2009, \apjl, 701, L1

\bibitem[{{DeForest} \& {Gurman}(1998)}]{1998ApJ...501L.217D}
{DeForest}, C.~E. \& {Gurman}, J.~B. 1998, \apjl, 501, L217

\bibitem[{{Gupta} {et~al.}(2010){Gupta}, {Banerjee}, {Teriaca}, {Imada}, \&
  {Solanki}}]{2010ApJ...718...11G}
{Gupta}, G.~R., {Banerjee}, D., {Teriaca}, L., {Imada}, S., \& {Solanki}, S.
  2010, \apj, 718, 11

\bibitem[{{Gupta} {et~al.}(2012){Gupta}, {Teriaca}, {Marsch}, {Solanki}, \&
  {Banerjee}}]{2012A&A...546A..93G}
{Gupta}, G.~R., {Teriaca}, L., {Marsch}, E., {Solanki}, S.~K., \& {Banerjee},
  D. 2012, \aap, 546, A93

\bibitem[{{Krishna Prasad} {et~al.}(2011){Krishna Prasad}, {Banerjee}, \&
  {Gupta}}]{2011A&A...528L...4K}
{Krishna Prasad}, S., {Banerjee}, D., \& {Gupta}, G.~R. 2011, \aap, 528, L4

\bibitem[{{Lemen} {et~al.}(2012){Lemen}, {Title}, {Akin}, {Boerner}, {Chou},
  {Drake}, {Duncan}, {Edwards}, {Friedlaender}, {Heyman}, {Hurlburt}, {Katz},
  {Kushner}, {Levay}, {Lindgren}, {Mathur}, {McFeaters}, {Mitchell}, {Rehse},
  {Schrijver}, {Springer}, {Stern}, {Tarbell}, {Wuelser}, {Wolfson}, {Yanari},
  {Bookbinder}, {Cheimets}, {Caldwell}, {Deluca}, {Gates}, {Golub}, {Park},
  {Podgorski}, {Bush}, {Scherrer}, {Gummin}, {Smith}, {Auker}, {Jerram},
  {Pool}, {Soufli}, {Windt}, {Beardsley}, {Clapp}, {Lang}, \&
  {Waltham}}]{2012SoPh..275...17L}
{Lemen}, J.~R., {Title}, A.~M., {Akin}, D.~J., {et~al.} 2012, \solphys, 275, 17

\bibitem[{{Mariska}(1992)}]{1992str..book.....M}
{Mariska}, J.~T. 1992, {The solar transition region} (New York: Cambridge
  University Press)

\bibitem[{{McIntosh} {et~al.}(2010){McIntosh}, {Innes}, {de Pontieu}, \&
  {Leamon}}]{2010A&A...510L...2M}
{McIntosh}, S.~W., {Innes}, D.~E., {de Pontieu}, B., \& {Leamon}, R.~J. 2010,
  \aap, 510, L2

\bibitem[{{Morgan} {et~al.}(2004){Morgan}, {Habbal}, \&
  {Li}}]{2004ApJ...605..521M}
{Morgan}, H., {Habbal}, S.~R., \& {Li}, X. 2004, \apj, 605, 521

\bibitem[{{Ofman} {et~al.}(2000){Ofman}, {Nakariakov}, \&
  {Sehgal}}]{2000ApJ...533.1071O}
{Ofman}, L., {Nakariakov}, V.~M., \& {Sehgal}, N. 2000, \apj, 533, 1071

\bibitem[{{Ofman} {et~al.}(2012){Ofman}, {Wang}, \&
  {Davila}}]{2012ApJ...754..111O}
{Ofman}, L., {Wang}, T.~J., \& {Davila}, J.~M. 2012, \apj, 754, 111

\bibitem[{{Pant} {et~al.}(2015){Pant}, {Dolla}, {Mazumder}, {Banerjee},
  {Krishna Prasad}, \& {Panditi}}]{2015ApJ...807...71P}
{Pant}, V., {Dolla}, L., {Mazumder}, R., {et~al.} 2015, \apj, 807, 71

\bibitem[{{Tian} {et~al.}(2011){Tian}, {McIntosh}, {Habbal}, \&
  {He}}]{2011ApJ...736..130T}
{Tian}, H., {McIntosh}, S.~W., {Habbal}, S.~R., \& {He}, J. 2011, \apj, 736,
  130

\bibitem[{{Wang} {et~al.}(2013){Wang}, {Ofman}, \&
  {Davila}}]{2013ApJ...775L..23W}
{Wang}, T., {Ofman}, L., \& {Davila}, J.~M. 2013, \apjl, 775, L23

\bibitem[{{Withbroe}(1983)}]{1983SoPh...89...77W}
{Withbroe}, G.~L. 1983, \solphys, 89, 77

\bibitem[{{Xia} {et~al.}(2005{\natexlab{a}}){Xia}, {Popescu}, {Chen}, \&
  {Doyle}}]{2005ESASP.592..575X}
{Xia}, L.~D., {Popescu}, M.~D., {Chen}, Y., \& {Doyle}, J.~G.
  2005{\natexlab{a}}, in ESA Special Publication, Vol. 592, Solar Wind 11/SOHO
  16, Connecting Sun and Heliosphere, ed. B.~{Fleck}, T.~H. {Zurbuchen}, \&
  H.~{Lacoste}, 575

\bibitem[{{Xia} {et~al.}(2005{\natexlab{b}}){Xia}, {Popescu}, {Doyle}, \&
  {Giannikakis}}]{2005A&A...438.1115X}
{Xia}, L.~D., {Popescu}, M.~D., {Doyle}, J.~G., \& {Giannikakis}, J.
  2005{\natexlab{b}}, \aap, 438, 1115

\bibitem[{{Yuan} {et~al.}(2014){Yuan}, {Sych}, {Reznikova}, \&
  {Nakariakov}}]{2014A&A...561A..19Y}
{Yuan}, D., {Sych}, R., {Reznikova}, V.~E., \& {Nakariakov}, V.~M. 2014, \aap,
  561, A19

\end{thebibliography}
\end{document}